\documentclass[%
 reprint, 
superscriptaddress, 
twocolumn,
nobibnotes,
 amsmath,amssymb,
 aps,
prx
]{revtex4-2}

\usepackage{siunitx}[=2021-04-09] 
\usepackage{spin-textures}
\usepackage{graphicx}
\usepackage{dcolumn}
\usepackage{tabularx}
\usepackage{physics}
\usepackage{enumerate}
\usepackage[utf8]{inputenc}
\usepackage{float}
\usepackage{amsmath,amssymb,amsthm,bm}
\usepackage{subfigure}
\usepackage{mathtools}
\usepackage[dvipsnames]{xcolor}
\usepackage{comment}

\begin{document}

\preprint{APS/123-QED}

\title{Topological analysis and experimental control of transformations of domain walls in magnetic cylindrical nanowires}

\author{L.~Álvaro-Gómez}
 \email{laura.alvaro@ucm.es}
 \affiliation{Univ. Grenoble Alpes, CNRS, CEA, Grenoble INP, SPINTEC, 38000 Grenoble, France.}
 \affiliation{IMDEA Nanociencia, Campus de Cantoblanco, 28049 Madrid, Spain.}
 \affiliation{Dpto. de Física de Materiales, Universidad Complutense de Madrid, 28040 Madrid, Spain.}

\author{J.~Hurst}
\affiliation{Univ. Grenoble Alpes, CNRS, CEA, Grenoble INP, SPINTEC, 38000 Grenoble, France.}

\author{S. Hegde}
 \affiliation{Univ. Grenoble Alpes, CNRS, CEA, Grenoble INP, SPINTEC, 38000 Grenoble, France.}
\affiliation{UM-DAE Centre for Excellence in Basic Sciences, University of Mumbai,  400098 Mumbai, India.}
\author{S.~Ruiz-Gómez}
\affiliation{Max Planck Institute for Chemical Physics of Solids, 01187 Dresden, Germany.}

\author{E. Pereiro}
\affiliation{Alba Synchrotron Light Facility, CELLS, 08290 Cerdanyola del Vallès, Barcelona, Spain.}

\author{L.~Aballe}
\affiliation{Alba Synchrotron Light Facility, CELLS, 08290 Cerdanyola del Vallès, Barcelona, Spain.}

\author{J.C~Toussaint}
\affiliation{Univ. Grenoble Alpes, CNRS, Institut Néel, 38000 Grenoble, France.}

\author{L.~Pérez}
 \affiliation{IMDEA Nanociencia, Campus de Cantoblanco, 28049 Madrid, Spain.}
 \affiliation{Dpto. de Física de Materiales, Universidad Complutense de Madrid, 28040 Madrid, Spain.}

\author{A.~Masseboeuf}
 \affiliation{Univ. Grenoble Alpes, CNRS, CEA, Grenoble INP, SPINTEC, 38000 Grenoble, France.}

\author{C.~Thirion}
\affiliation{Univ. Grenoble Alpes, CNRS, Institut Néel, 38000 Grenoble, France.}

\author{O.~Fruchart}
 \affiliation{Univ. Grenoble Alpes, CNRS, CEA, Grenoble INP, SPINTEC, 38000 Grenoble, France.}

\author{D.~Gusakova}
\email{daria.gusakova@cea.fr}
 \affiliation{Univ. Grenoble Alpes, CNRS, CEA, Grenoble INP, SPINTEC, 38000 Grenoble, France.}

\date{\today}

\begin{abstract}
Topology is a powerful tool for categorizing magnetization textures by defining a topological index in both 2D systems, such as thin films or curved surfaces, and in 3D bulk systems. In the emerging field of 3D nanomagnetism,  both volume and  surface topological numbers must be considered, requiring the identification of a proper global topological invariant to support categorization. Here, we consider domain walls in cylindrical nanowires as an excellent playground for 3D nanomagnetic systems, excited by a charge current, that generates an \OErsted field. We first provide experimental evidence of previously-unreported domain-wall transformations of topology occurring at the nanosecond time scale. We investigate these transformations with micromagnetic simulations, tracking both bulk and surface topological signatures. We demonstrate a topological invariant combining both signatures, while the topological charge varies from bulk to surface during the dynamics. The experimental change of topology is reproduced when the pulse duration matches the time scale of the internal transformations of the wall, and the current is switched off before the transformation is complete. We expect that the topological categorization and dynamical exploitation apply to any 3D nanomagnetic system.

\end{abstract}

\maketitle


\section{\label{sec-intro}Introduction}

Topology has been for long a helpful approach to categorize nematic, magnetic and other physical textures \cite{bib-MER1979}. In magnetic thin films and nanostructures, this classification includes domain walls~(DWs), vortices/antivortices, and skyrmions/antiskyrmions. The emerging field of 3D nanomagnetism allows the existence of additional  magnetization textures, including Bloch points, vortex loops, hopfions, and bobbers, among others \cite{bib-STR2021}. Topological invariants allow the categorization of vector fields corresponding to each magnetic texture from the viewpoint of the possibility of  continuous deformation. In 3D nanomagnetic finite-size systems, in contrast to flat thin films, the calculation of topological invariants involves distinct mathematical formalisms for the volume and for the surface, making analysis of a system as a whole more complex. In particular, this raises the question how topological invariants of two sub-spaces are connected together, and whether they allow to follow any type of transformations of the magnetization texture.

DWs in cylindrical nanowires made of magnetically soft materials represent a textbook case for investigating 3D nanomagnetism in finite size systems. Indeed,  nanowires exhibit a geometrical rotation invariance and under some conditions, DWs  may be considered as nearly one dimensional, reducing complexity. Still, when the diameter exceeds approximately seven times the dipolar exchange length (\ie up to a few tens of nanometers), complex 3D magnetization textures may develop \cite{bib-FRU2015b}, involving 3D nanostructure features. The moderate dimensions of these structures allow for accurate micromagnetic simulations, and experiments are feasible using electrodeposition as a versatile growth method for high quality nanomaterials \cite{bib-SOU2014}. Two types of DWs topologies exist in such wires, both energetically metastable and with very similar energy: the Bloch-Point-Wall (BPW) and the Transverse-Vortex-Wall (TVW) \cite{bib-FOR2002,bib-HER2002a,bib-THI2006,bib-YAN2010, bib-FRU2015b}. Experimental observations have  demonstrated that these DWs can dynamically change  topology under the application of a sufficiently large stimulus, either a longitudinal applied magnetic field \cite{bib-FRU2019} or a charge current inducing an \OErsted field  \cite{bib-FRU2019b}. Simulations  conducted under quasi-static conditions (with high damping $\alpha=1$, and moderate driving force) revealed the basics of the underlying microscopic phenomenon, which involves the annihilation and/or the  nucleation of a Bloch point, and the existence of a pair of surface vortex/antivortex (V/AV). Notably, even when the initial and final states have the same topology, such as the switching of sign of circulation of a BPW, intermediate steps may not. An empirical conservation rule was proposed that allows to describe the magnetization texture at any time during a transformation, linking 2D and 3D topological numbers \cite{bib-FRU2021}. In this report, we provide the experimental proof of the topological transformations occurring during the BPW circulation switching, exhibiting a TVW as final state for nanosecond-short \OErsted field pulses, reminiscent of the transient change of topology. We demonstrate that it is possible to predictably convert the topology of a DW from BPW to TVW or vice-versa, depending on the duration of the current pulse. We  also theoretically generalize the topological rule to transformations involving a larger number of Bloch points and surface vortices and antivortices, which with micromagnetic simulations we predict to occur for large stimuli or lower and thus realistic damping conditions.

\section{\label{sec:methods}Methods}
\subsection{\label{sec:exp-methods}Experimental methods}
Permalloy (Fe\(_{20}\)Ni\(_{80}\)) cylindrical nanowires were synthesized using template-assisted electrochemical deposition. Nanoporous anodic aluminum oxide (AAO) templates were prepared by hard anodization of Al disks (Goodfellow, \SI{99.999}{\%} in purity) in a water-based solution of oxalic acid (\SI{0.3}{M}) and ethanol (\SI{0.9}{M}), applying \SI{140}{V} of anodization voltage at \SIrange{0}{1}{\celsius} for \SI{2.5}{h}. The remaining Al was etched with an aqueous solution of CuCl$_2$ (\SI{0.74}{M}) and HCl (\SI{3.25}{M}), the oxide barrier was removed and the pores opened to a final diameter of \SI{120}{nm} with H$_3$PO$_4$ (5 {\% vol.}).

The growth conditions were similar to those described in \cite{bib-RUI2018}. After the electrochemical growth, the AAO templates were dissolved in H\(_{3}\)PO\(_{4}\)(\SI{0.4}{M}) and H\(_{2}\)CrO\(_{4}\)(\SI{0.2}{M}). Then, the nanowires were dispersed on \SI{20}{nm} thick  Si\(_{3}\)N\(_{4}\) windows to allow for transmission microscopy. Nanowires were contacted electrically using laser lithography in order to allow the injection of high-frequency current pulses, revealing resistivity values  of approximately \SI{20}{\micro\ohm\centi\metre}. In order to enhance thermal dissipation a \SI{20}{nm} thick conformal layer of Al$_2$O$_3$ was deposited on the device with Atomic Layer Deposition (ALD). This broadens the contact area between the wire and its supporting surface, allowing for more heat transfer.  Additionally, to enhance thermal dissipation across the Si\(_{3}\)N\(_{4}\) membrane, a \SI{100}{nm} layer of Al was deposited on the backside of the substrate, while still allowing for transmission microscopy.

The wires were imaged using X-ray Magnetic Circular Dichroism (XMCD) coupled to Transmission X-ray Microscopy (TXM) \cite{Sorrentino2015} at the MISTRAL beamline of the ALBA Synchrotron. The photon energy was set to the Fe L$_{3}$ absorption edge, and the X-ray incidence was set 10° off the normal to the nanowire axis in order to be primarily sensitive to the azimuthal magnetization component, while allowing for a weak sensitivity to the longitudinal component of magnetization. The selected rotation angle represents a good compromise, as an incidence angle of 0° permits observation of only azimuthal magnetization, while angles exceeding 20° result in the axial magnetic contrast obscuring the azimuthal component. The calculated XMCD images stands for $ (\mathrm{I}_{+} -  \mathrm{I}_{+}) / (\mathrm{I}_{-}+ \mathrm{I}_{+}) $, where I$_{+}$ and  I$_{- }$ correspond to imaging  a region of the sample with right or left circularly polarized X-rays, respectively.

Series of about 64 images of \SI{1}{\second} of duration per X-ray polarization were acquired and drift-corrected prior to averaging in order to increase the signal-to-noise ratio without compromising spatial resolution. If not initially present, DWs were nucleated by applying, when on the TXM stage, a \SI{10}{ns}-long pulse of current density with an amplitude of \SI{1e12}{\ampere\per\meter\squared}. At such a relatively high and prolonged pulse duration, the sample demagnetizes to a new initial state including domain walls.

\subsection{\label{sec:theormodel}Micromagnetics}

Micromagnetic simulations were conducted with the \textit{feeLLGood} software \cite{bib-ALO2006,bib-ALO2012,bib-KRI2014,bib-FEE}, a home-made micromagnetic code based on finite-element methods. This code enables the accurate modeling of curved systems by discretizing the system into a mesh made of tetrahedrons. The \textit{feeLLGood} software solves the Landau-Lifshitz-Gilbert (LLG)-Slonczewski equation, which takes into account the effect of the spin transfer torque (STT), if relevant \cite{bib-ZHA2004a, bib-THI2005, bib-STU2016}. Here, we focus on the key role of the current-induced \OErsted field which was shown to be the driving phenomenon for DW dynamics in cylindrical nanowires \cite{these-arnaud}. Thus, we intentionally disregard spin-transfer torques, so that the time evolution of the unit magnetization vector $\vect{m}$ reads:
\begin{equation}
\dot{\vect{m}} =
-\gamma_0 \vect{m}\crossproduct\vect{\Heff}  +  \alpha\vect{m}\crossproduct \dot{\vect{m}}
\label{eq:LLGLong}
\end{equation}
where $\gamma_0 = \mu_0 \left| \gamma \right|$ \cite{bib-LAN1935} is the gyromagnetic ratio, $\alpha$ is the Gilbert damping
coefficient, and $\vect{H_\mathrm{eff}}$ is the effective field acting on the system (sum of the \OErsted, anisotropy, exchange and magnetostatic  fields). The expression for the \OErsted field in cylindrical coordinates reads $\mu_0 \vectHOE=\mu_0 J\rho \unitvectvarphi/2$, with $J$ being the current density.

\begin{figure}[ht]
\centering \includegraphics[width=0.8\linewidth]{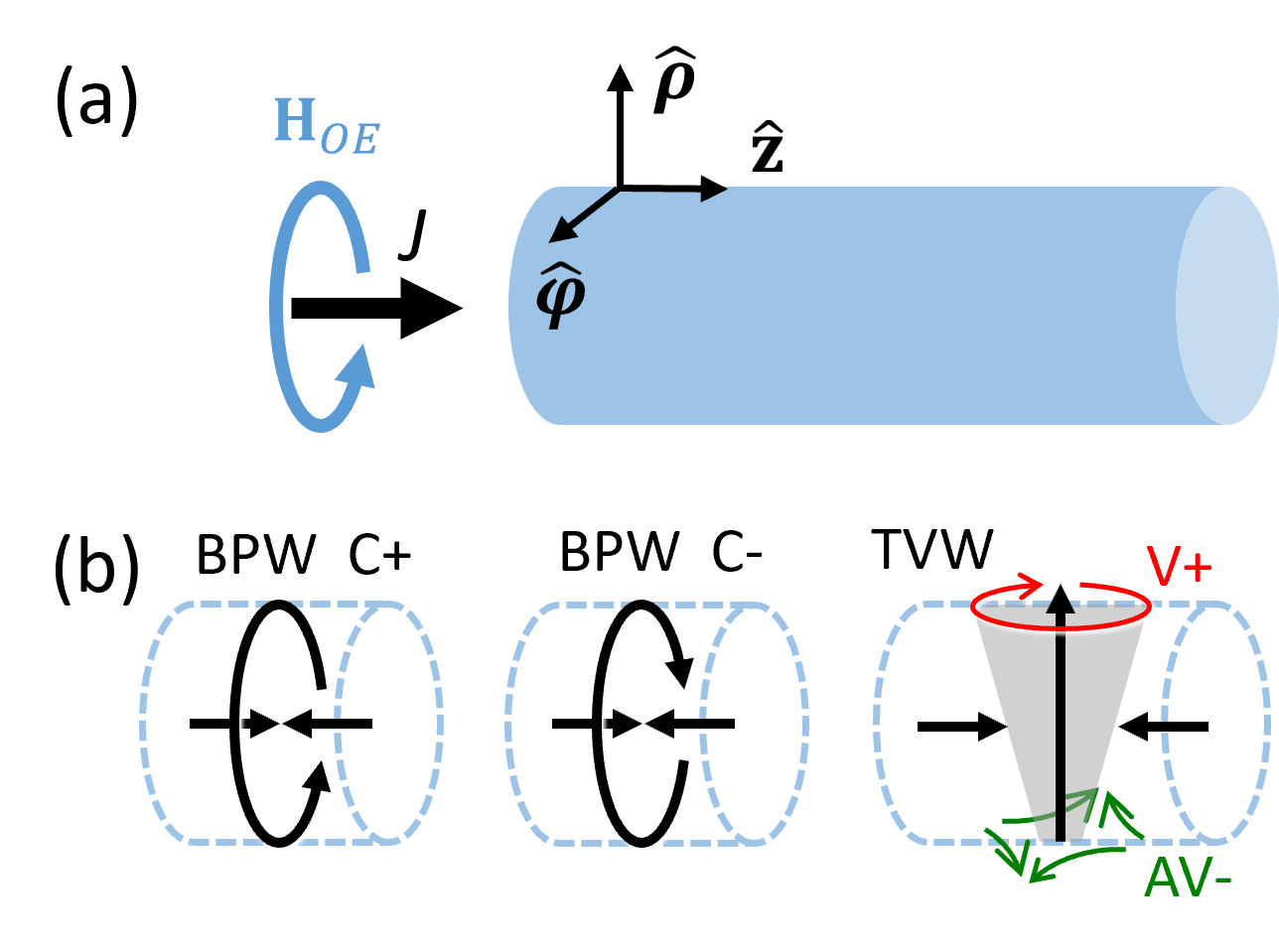}
\caption{\label{fig:geometry-axis}(a) Schematics of the geometry of the cylindrical nanowire considered, and the current-induced \OErsted field (b)~Sketch of head-to-head BPW texture with positive (C+) and negative (C-) circulation, defined with respect to $\unitvectz$, and head-to-head TVW texture. Light arrows sketch the direction of magnetization around the surface vortex~(resp. antivortex), here with positive V+~(resp. negative AV-) polarity, reflecting the sign of radial magnetization at the core of the object.}
\end{figure}

We consider cylindrical magnetic nanowires homogeneously made of Permalloy (Fe\(_{20}\)Ni\(_{80}\)). A schematic of this system, along with cylindrical coordinate axes, is depicted in \subfigref{fig:geometry-axis}{a}. We consider a wire diameter of \SI{90}{nm}, for which both types of DWs have been predicted by simulation to exist at rest \cite{bib-FOR2002,bib-HER2002a} and later confirmed experimentally \cite{bib-BIZ2013,bib-FRU2014}. This value is smaller than the experimental diameter, which allows significant gains in computing time especially as the wall width roughly scales like the square of the diameter in this regime \cite{bib-FRU2015b}. We checked that this does not change the qualitative conclusions.

Surface charges ($\sigma = -M_{\mathrm{s}} \vect{m} \cdot \vect{n}$) were removed numerically at both wire ends to cancel out magnetostatics at these ends and mimic an infinitely-long wire. The following material parameters were used: spontaneous magnetization $\Ms = \SI{8e5}{A/m}$, zero magnetocrystalline anisotropy, and exchange stiffness $A = \SI{1.3e-11}{J/m}$. The resulting dipolar exchange length $\DipolarExchangeLength = \sqrt{\frac{2A}{\muZero\Ms^2}}$ is $\SI{5.6}{nm}$. The mesh size used was \SI{4}{nm}. Finally, we considered two damping coefficients, either a moderate damping situation with $\alpha = 0.1$ or a high damping situation with $\alpha = 1$. Magnetization dynamics is driven by a homogeneous charge current flowing along the wire axis $z$, resulting in an analytically-calculated radially-dependent \OErsted field exerting a torque on magnetization. A step-like current density is applied to the system, with the starting point being the relaxed configuration of a BPW at rest, \ie, with no current. While there exist four degenerate instances of BPWs at rest, differing in head-to-head versus tail-to-tail domains, and the sign of circulation, the switching mechanisms are equivalent under time or space symmetries. We can therefore focus on one type of initial state, without loss of generality. In this manuscript, we consider head-to-head domains and positive initial circulation, defined with respect to $\unitvectz$, subjected to an \OErsted field pulse with negative circulation. The DW width may be followed throughout the switching process by monitoring the Thiele wall parameter, which is defined as:
\begin{equation}
\Delta_{\mathrm{T}} = \frac{2S}{\int\left(\partial_{\mathrm{z}}\vect{m}\right)^2\mathrm{d}V},
\label{eq:ThieleWidth}
\end{equation}
where $S$ is the cross-sectional area of the nanowire.

\subsection{\label{sec:singularities}Tracking of surface vortices/antivortices}

The BPW is characterized by magnetization curling around the wire axis (\subfigref{fig:geometry-axis}{b}). Consequently, the magnetization remains largely parallel to the wire surface at all places, thereby minimizing surface magnetic charges. As these surface boundary conditions do not allow a continuous vector field to map the bulk, a magnetic singularity called the Bloch point is required to exist within the bulk \cite{bib-FEL1965, bib-DOE1968}. In practice and at rest, the Bloch point is located on the axis of the wire, at the center of the wall. The TVW is characterized by a component of magnetization transverse to the wire axis (\subfigref{fig:geometry-axis}{b}), with two loci having full radial magnetization at the wire surface, one outward and one inward. When the wire diameter exceeds approximately seven times $\DipolarExchangeLength$ \cite{bib-FRU2015b}, magnetization tends to curl around these loci, reminiscent of vortex/antivortex patterns. The vortex core can point either outward or inward (along the axis perpendicular to the plane of the vortex), giving it a "positive" or "negative" polarity, respectively. While Bloch points are micromagnetic singularities \cite{bib-FEL1965,bib-DOE1968}, vortices and antivortices are not, as the vector field for magnetization in these regions remains continuous. We will refer to any of these two as V/AV in the following.

The occurrence of one or the other type of topological objects uniquely defines the topology of a given DW. Therefore, it is convenient to monitor them to analyze DW dynamics and possible transformations, which  may involve a larger variety of situations in the transient state than at rest.  To do so, we use a in-house post-processing tool for Bloch point and V/AV tracking, developed specifically for tetrahedron-based finite elements and inspired by vector fields metrics and critical points analysis applied to computer visualization \cite{Hofmann2018}. The output lists all topological objects, along with their type and coordinates. The classification of topological objects consists in calculating the Jacobian matrix near the critical point and analyzing  its eigenvalues. This mathematical procedure is implemented distinctly in two and three dimensions. Critical points are sorted into several categories according to the sign of real and imaginary parts of the eigenvalues. On the wire surface, we obtain a spiral-type solution for the vortex and a saddle-type solution for the antivortex. In the volume of the wire the head-to-head Bloch Point corresponds to the spiral-saddle-in-type solution, whereas the tail-to-tail Bloch Point corresponds to the spiral-saddle-out-type solution.

\section{Experimental proof of the change of topology during dynamics}\label{sec-experiments}

\begin{figure}[!ht]
\centering \includegraphics[width=\linewidth]{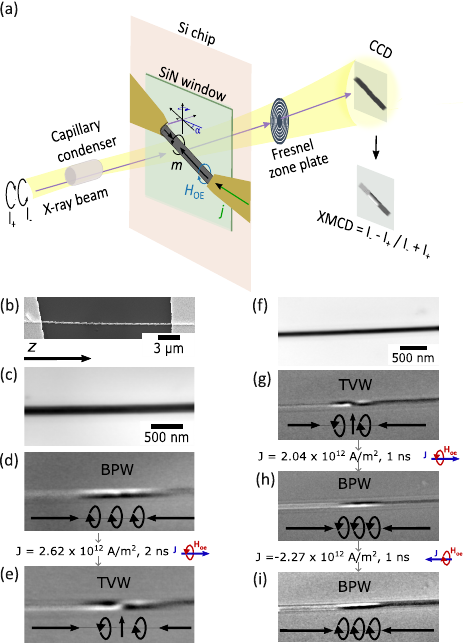}
\caption{Transmission X-ray microscopy (TXM) imaging of current-induced domain-wall transformations in a \SI{120}{nm}-diameter Permalloy nanowire. (a) Schematic of a TXM setup for imaging an electrically-contacted nanowire exhibiting the magnetic state of a BPW. (b) Electron microscopy image of an electrically contacted Permalloy nanowire. (c) and  (f)  TXM images at the Fe L$_{3}$ edge of different regions of the nanowire.  The corresponding initial magnetic configurations are shown in the XMCD images in (d) and (g). The black arrows indicate the direction of magnetization. Subsequent XMCD images, after the application of the indicated current pulse, show  (e) BPW to TVW transformation, (h) TVW to BPW transformation and (i) BPW circulation switching.  }
\label{fig:exp-BPW-to-TVW-OE-homogeneous-wire}
\end{figure}

We experimentally characterized the three-dimensional magnetic texture by performing TXM-XMCD imaging of electrically-contacted nanowires. An illustrative schematic of  the setup  is shown in \subfigref{fig:exp-BPW-to-TVW-OE-homogeneous-wire}{a}, where a region of the nanowire is illuminated with right or left circularly polarized X-rays. In this schematic, the magnetic configuration displays a BPW, which we expect to result in pronounced dark and bright bipolar XMCD contrast across the wire at this location, as the magnetization is nearly parallel or anti-parallel to the X-ray beam on the two sides of the wire. Due to the $\alpha=$\SI{10}{\degree} rotation of the sample holder, we also expect that the oppositely-aligned axial magnetic domains are revealed by a bright and dark grey contrast. \subfigref{fig:exp-BPW-to-TVW-OE-homogeneous-wire}{b} displays an electron microscopy image of a Permalloy nanowire with a diameter of  \SI{120}{nm}, featuring a gold pad at each termination.

Magnetic domain walls were obtained by demagnetizing the nanowire with a relatively large current pulse. \subfigref{fig:exp-BPW-to-TVW-OE-homogeneous-wire}{c} displays a TXM image at the Fe L$_3$ edge of a region of Permalloy nanowire. The corresponding XMCD image (\subfigref{fig:exp-BPW-to-TVW-OE-homogeneous-wire}{d}) reveals a dark and bright bipolar contrast, characteristic of the curling magnetization in a BPW. The light dark and bright magnetic contrast observed to the right and left sides, respectively, reveals the existence of head-to-head axial domains, thanks to the $\SI{10}{\degree}$ tilt of the beam of X-rays. The combination of these two contrasts is a unique signature for a BPW with negative circulation C$-$ with respect to $\unitvectz$. Thanks to the opposite contrast arising from the neighboring domains, we can here dismiss the possibility of the central curling being due to an azimuthal domain, sometimes found in nanowires above a certain diameter approximately above \SI{150}{nm} \cite{bib-BRA2017,bib-RUI2018}.

In previous investigations, the switching of the circulation of BPWs driven by the \OErsted field was conducted using current pulses lasting around \SI{15}{ns} \cite{bib-FRU2019b}. In the present study, we considered shorter pulse duration, typically \SIrange{1}{2}{ns}, of the order of the duration of the magnetization dynamics predicted in micromagnetic simulations \cite{bib-FRU2021}. This results in a change in behavior in certain events. \subfigref{fig:exp-BPW-to-TVW-OE-homogeneous-wire}{e} displays the qualitative XMCD image obtained after applying a \SI{2}{ns} current pulse along $\unitvectz$ with an amplitude of \SI{2.62e12}{\ampere\per\meter\squared}, \ie, giving rise to an \OErsted field antiparallel to the initial BPW circulation. The observed magnetic contrast indicates the transformation of the BPW into a TVW.  This current-induced transformation had not been reported before, highlighting the significant impact that a very short pulse duration has on the dynamics. \subfigref{fig:exp-BPW-to-TVW-OE-homogeneous-wire}{f} shows a TXM image at the Fe L$_3$ edge of another region of the nanowire. Its corresponding XMCD image reveals a TVW configuration as the initial state (\subfigref{fig:exp-BPW-to-TVW-OE-homogeneous-wire}{g}). Upon the application of a current pulse of similar characteristics, the final state exhibits a BPW configuration (\subfigref{fig:exp-BPW-to-TVW-OE-homogeneous-wire}{h}).  Finally, following the application of another pulse coming with an \OErsted field with antiparallel circulation, the BPW reverses its circulation, as previously reported \cite{bib-FRU2019b, bib-FRU2021}. These types of domain wall transformations were also observed when lower amplitude pulses were applied, as low as \SI{0.5e12}{\ampere\per\meter\squared}, however with longer duration, up to \SI{5}{ns}.

These experimental observations demonstrate that the transformations from a BPW into a BPW of reversed circulation and from a TVW into a BPW also take place, consistent with our theoretical predictions \cite{bib-FRU2021}. In addition, we also evidenced that a BPW can transform into a TVW, when a sufficiently-short current pulse with an anti-parallel \OErsted field is applied. Although the transformations are not always reproducible under similar stimulus - possibly due to  contributions from STT and/or Joule heating   - the possibility to end up in a TVW is abundant and thus unambiguously proven. The fact that a BPW can either transform to a TVW or reverse its circulation under a similar current pulse, suggests that the experimental pulse duration is comparable with the timescale of the dynamics.  Consequently, in the subsequent sections, we investigate through micromagnetic simulations the mechanism of DW transformations, with a special emphasis on the influence  of the pulse duration.

\section{\label{sec:bpw-switching}{Micromagnetic simulation of BPW circulation switching}}

\begin{figure*}
\centering\includegraphics[width=\textwidth]{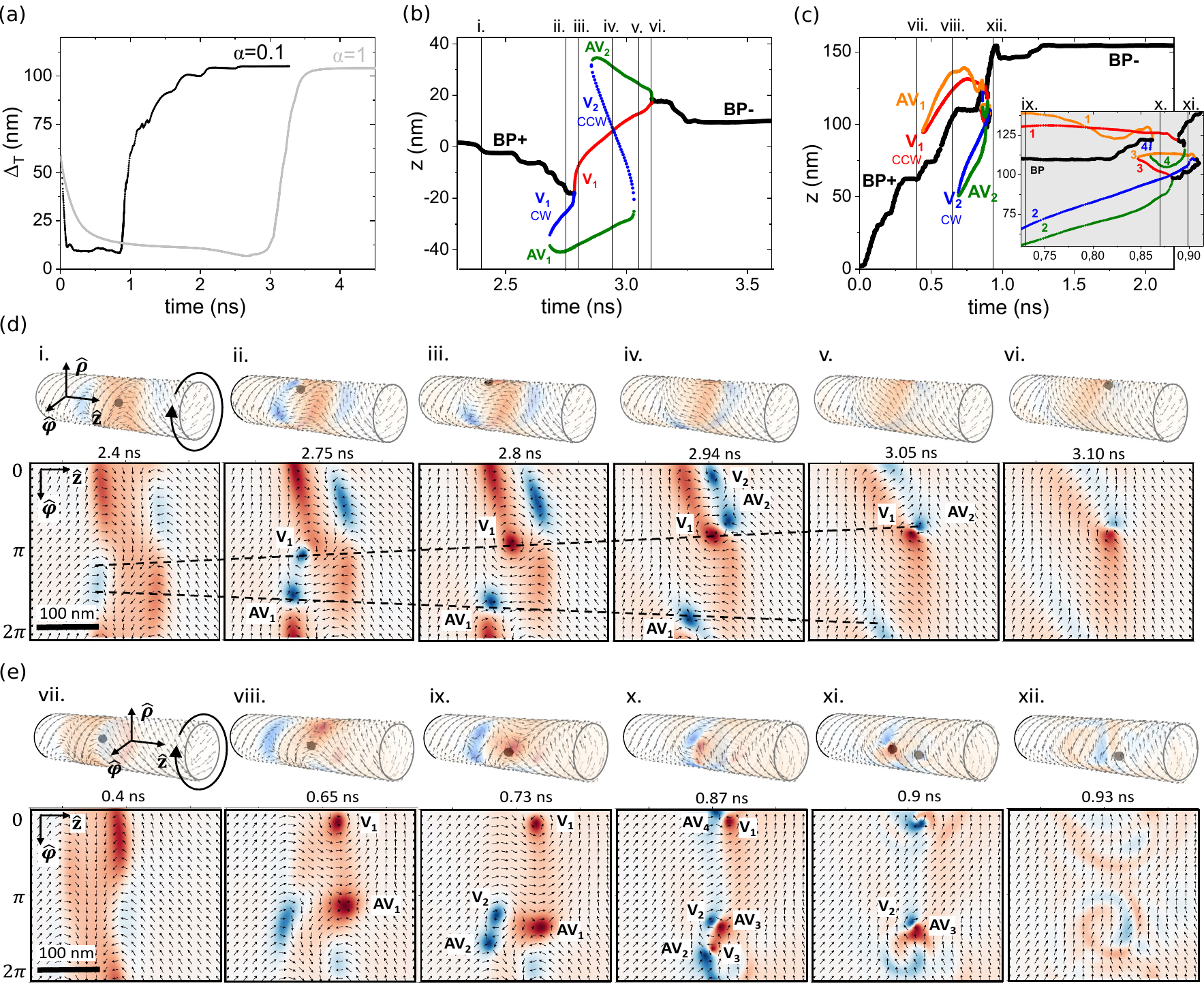}
\caption{Switching of circulation of a head-to-head BPW in a \SI{90}{nm}-diameter homogeneous Permalloy nanowire, for damping values $\alpha=1$ and $\alpha=0.1$. A pulse of current density of $\SI{-1.50e12}{\ampere\per\meter\squared}$ is applied in a step-like manner at $t=0$, resulting in an \OErsted field anti-parallel to the initial BPW circulation, with a magnitude of  \SI{42.5}{\milli\tesla} at the surface of the wire. (a) Domain wall width $\Delta_{\tau}$ as a function of time for both damping parameters. (b) and (c): $z$-axis trace of the topological objects versus time for $\alpha =1$ and $\alpha = 0.1$, respectively. The colors indicate different topological features: black for  Bloch points, red (blue) for vortices with positive (negative) polarity, orange (green) for anti-vortices with positive (negative) polarity. (d) and (e): Snapshots of the magnetic texture at selected times, for $\alpha = 1$ and $\alpha = 0.1$, respectively. The selected times are highlighted in (b) and (c), respectively. The top rows show 3-dimensional external surface views of the nanowire, while the bottom rows despict its unrolled 2-dimensional ($\unitvectz$ - $\unitvectvarphi$) cylinder surface, viewed from the outside. The color map represents the radial magnetization component, where $m_{\rho}=1$ is red and $m_{\rho}=-1$ is blue. Dashed lines highlight the trajectories of $\mathrm{V}_1$ and $\mathrm{AV}_1$.}
\label{fig:BPW-switching}
\end{figure*}

\begin{figure}[t]
\centering\includegraphics[width=0.8\linewidth]{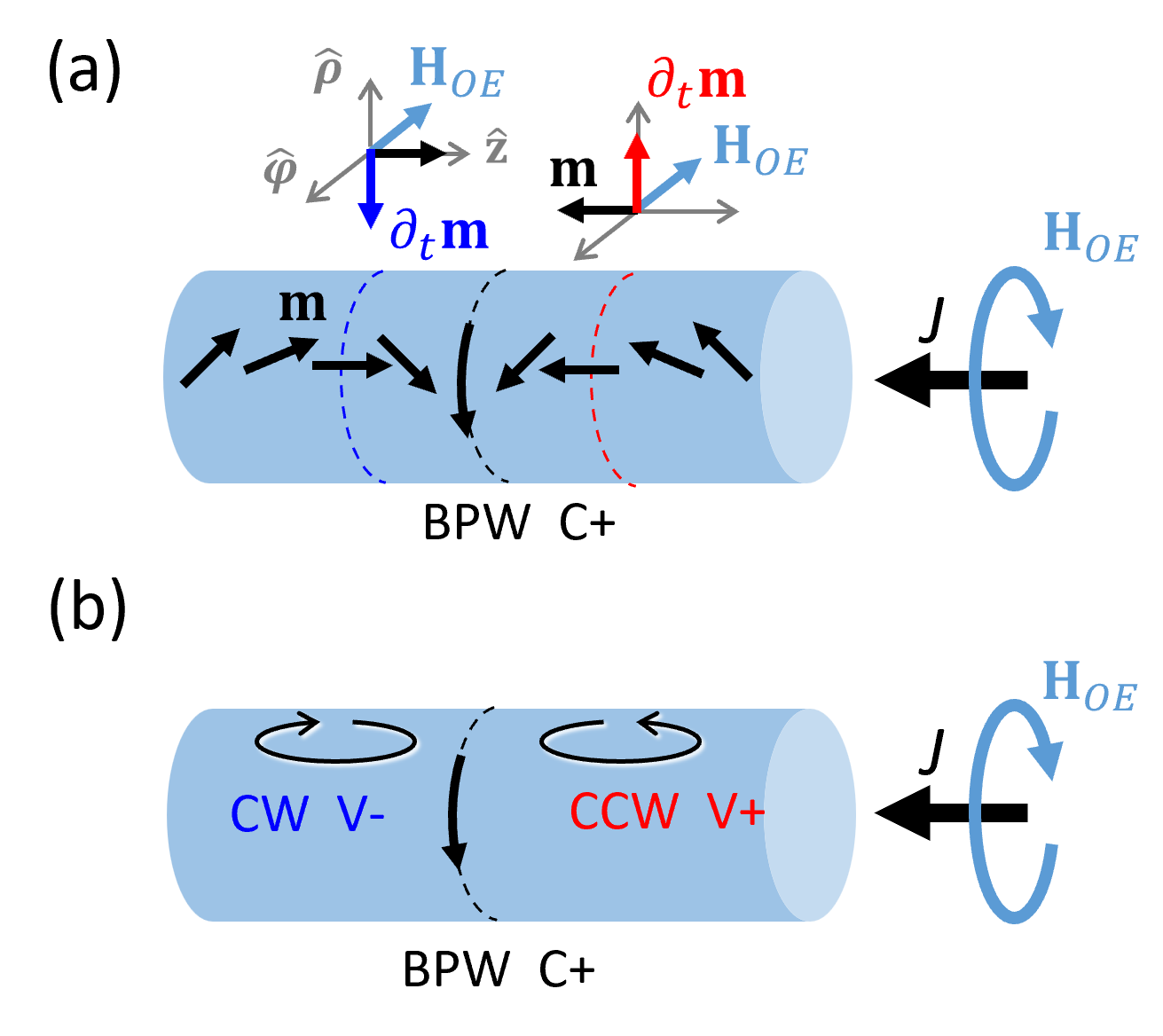}
\caption{  Identification of the type of the first vortex nucleated based on the sign of the \OErsted field radial torque during BPW circulation switching from C+ to C-. (a) A simplified illustration of the \OErsted field induced torque ($\partial_{t}\vect{m}$) along the $m_{\varphi}=0$ isolines (blue and red dashed lines) prior to the switching process. To the left of the BPW center (black dashed line), the magnetization aligns towards negative radial values ($m_{\rho}<0$)  while it aligns towards positive radial values  ($m_{\rho}>0$)  on the right side. (b) Illustration of the vortex in-plane orientation (clockwise or counterclockwise) while switching. }
\label{fig:V-AV-polarity}
\end{figure}

The BPW switching mechanism was previously studied by micromagnetic simulations under quasistatic conditions to evidence the basics of the mechanism and a minimum energy path \cite{bib-FRU2021}: damping was nonphysically-high with $\alpha=1$, and the applied current density was only slightly above the switching threshold. Here, we explore a broader range of situations. First, we consider both high~($\alpha=1$) and moderate damping~($\alpha=0.1$), and a larger current density amplitude: $\SI{1.50e12}{\ampere\per\meter\squared}$, well above the threshold for switching  $J_{\mathrm{c}}= \SI{0.80e12}{\ampere\per\meter\squared}$. The applied current density amplitude of $\SI{1.50e12}{\ampere\per\meter\squared}$ corresponds to an \OErsted field magnitude of \SI{42.5}{\milli\tesla} at the periphery of the wire of \SI{90}{nm} diameter. We recall that the diameter considered  in the simulations is lower than that in the experiments, for the sake of the computational time.

\figref{fig:BPW-switching} illustrates the switching process of the circulation of the BPW for both high and moderate damping, highlighting the DW width evolution in time, the longitudinal position of topological objects detected, and surface magnetization maps~(see \figref{fig:BPW-switching} caption). Starting from the configuration at rest, the circulation switching can be decomposed into four main steps. \textit{Phase 1}: The magnetization in the domains undergoes a torque from the \OErsted field, which rapidly tilts the magnetization towards the azimuthal direction $\unitvectvarphi$, with a final angle dependent on the amplitude of the \OErsted field and the radial position. At the wire surface, this increases the angle of the DW above $\SI{180}{\degree}$. Note that the central part of the BPW remains mostly unaffected as little torque is exerted, since the magnetization is largely antiparallel to the \OErsted field. The DW rapidly shrinks under the action of the \OErsted field (\figref{fig:BPW-switching}{a}). \textit{Phase 2}: At the two transition regions of the DW, largely displaying axial magnetization, the torque arising from the \OErsted field leads to a gradual increase of the radial component of magnetization. A breaking of the rotational symmetry is developing over time, the radial component of magnetization becoming azimuth-dependant. This process takes time, with no significant change in the DW width. \textit{Phase 3}: A pair of surface V/AV nucleates when full radial magnetization is reached at a given locus \cite{bib-HER2007}. This is followed by a complex dynamics involving their motion around the wire periphery. 
Other processes involving change of polarity, annihilation or further nucleation may occur, which will be discussed in more detail hereafter.  During most of this sequence, the DW width remains small. \textit{Phase 4}: The topology of a BPW with circulation parallel to the \OErsted field is achieved and remains. The DW then rapidly expands, to reach a width larger than at rest.

Phases 1 and 4 are qualitatively robust insensitive to the value of current density and damping. Quantitatively, the dynamics is faster for larger current densities and lower damping, although ringing effects arise in the latter case. In the following, we focus on phases 2 and 3, which depend on these two parameters. Consistent with the case of high damping and close-to-threshold current density \cite{bib-FRU2021}, \subfigref{fig:BPW-switching}{d} and \subfigref{fig:BPW-switching}{e} show that surface V/AV appear in pairs and with the same polarity. This is a continuous process, without the need for creating or annihilating a Bloch point, as shown in \subfigref{fig:BPW-switching}{b-c}. However, while in Ref.\cite{bib-FRU2021} only one pair of V/AV is nucleated, here another one appears at $t=\SI{2.85}{\nano\second}$ and $z=\SI{33}{nm}$ for $\alpha=1$, and three appear for $\alpha=0.1$~(at \SI{0.69}{\nano\second}, \SI{0.85}{\nano\second} and \SI{0.86}{\nano\second}). This illustrates that new pathways arise for larger driving force or lower damping, which is a usual micromagnetic phenomenon, as the torque deriving from the driving force allows the magnetization texture to explore a larger phase space. 

Despite the variety and apparent complexity of behaviors observed, the processes involving  surface and volume objects always obey the same rules, whatever their total number is:
   \begin{enumerate}[a.]
     \item Surface V/AV nucleate in the form of a pair of vortex plus antivortex with the same polarity. Conversely, two such objects, \ie, with the same polarity, may annihilate continuously.
     \item A Bloch point annihilates upon reaching the surface \ie, it leaves the volume. It happens to reach the surface channeling along the core of a V or AV, such as at $t=\SI{2.78}{\nano\second}$ for $\alpha=1$. The channeling effect can be explained by the fact that a V or AV texture is found around any Bloch point, so that the two magnetization textures fit and may overlap. Upon the Bloch Point annihilation the polarity of the surface V/AV switches, consistent with the head-to-head or tail-to-tail nature of the Bloch point and the fact that all $4\pi$ orientations of the magnetization vector must be found in its vicinity \cite{bib-FRU2021}.
     \item The merging of a V and AV with opposite polarity gives rise to the nucleation of a new Bloch point. This process may indeed be decomposed in the nucleation of Bloch point and V/AV core switching time reversal symmetry of rule~b, followed by the continuous annihilation of V/AV with the same polarity~(rule~a). Once nucleated, the Bloch point then moves towards the inside of the wire.
     \item A pair of V and AV that nucleated together are not bound together, they may recombine with other AV and V.
     \item Two V or two AV do not nucleate nor annihilate together.

   \end{enumerate}

The rules stated above are consistent with microscopic processes reported in other contexts, such as the switching of polarity of a vortex subject to an external axial magnetic field \cite{bib-THI2003} or dynamically when exceeding a threshold velocity for a vortex \cite{bib-VAN2006} or antivortex \cite{Gliga2008}, the latter two processes involving also the continuous nucleation of a V/AV pair with identical polarity.

Let us provide a more quantitative insight about the first V/AV pair nucleated during the switching process. As observed in \subfigref{fig:BPW-switching}{d}, the first set of V/AV appears on the left side of the DW, with negative polarity, while in \subfigref{fig:BPW-switching}{e}, it appears on the right side, with positive polarity. As a matter of fact, even though the BPW is not chiral at rest, it becomes chiral during its dynamics \cite{bib-THI2006,bib-YAN2012}. \subfigref{fig:V-AV-polarity}{a} illustrates this by sketching the radial component of the torque induced on magnetization due to the \OErsted field with negative circulation, on both sides of the BPW C+: $\mathrm{d}m_{\mathrm{\rho}}/\mathrm{d}t$ is negative on the left side and positive on the right side. This is consistent with the different polarity for the first V/AV pair, depending on the side of its nucleation. Thus, it is probable that a V/AV pair that nucleates on the left side of a head-to-head BPW has negative polarity, while one that nucleates on the right side has positive polarity, resulting from the sign of the radial component of the Oersted field induced torque. The understanding of why the polarity of the first V/AV pair depends on damping is more speculative.  One interpretation could be that in the quasi-static case with $\alpha=1$, flux closure and  minimization of the dipolar energy are promoted since the associated charge of a head-to-head BPW is positive, which may facilitate the nucleation or neighboring regions with $m_\rho<0$. Conversely, for $\alpha=0.1$, precessional effects are more important, which may lead to the rapid increase of $m_\rho$ on the right side of the DW, quickly adding up to the pre-existing radial component. However, it is important to highlight that in this study thermal stochastic effects are not included, leading to a deterministic dynamics.

Regarding the in-plane circulation of vortices with respect to the outer normal -- clock-wise or counter-clock-wise --, it is deterministically related to the BPW circulation and whether they nucleate to the left or to right as illustrated in \subfigref{fig:V-AV-polarity}{b}. Vortices cannot change their initial in-plane circulation during their subsequent dynamics. Instead of this, they drift to the opposite side of the BPW \ie, left versus right, to fit its final circulation after the switching. This is clearly seen in \subfigref{fig:BPW-switching}{d} for $V_1$ and $V_2$.

Let us comment on the discrepancies between the experimental and simulated DW widths, which are about \SI{800}{nm} in \figref{fig:exp-BPW-to-TVW-OE-homogeneous-wire}, versus \SI{100}{nm} in \subfigref{fig:BPW-switching}{a}. The value estimated from experiments relies on the dichroic contrast, which is related to the global magnetization texture. Additionally, in TXM, we are sensitive mostly to the peripheral magnetization where magnetization is colinear with the X-ray beam. Micromagnetic simulations have shown that this is the region where the DW reaches its largest extension \cite{bib-FRU2015b}. Consequently, the apparent experimental width corresponds to the full width of the DW, being \SI{800}{nm}  consistent with the distribution of magnetization found by micromagnetic simulations  \cite{bib-FRU2015b}.  The DW width computed from micromagnetic simulations is based  on the Thiele definition (\eqnref{eq:ThieleWidth}). This definition is the inverse of the average squared gradient of the magnetization, giving more importance to regions with a larger gradient. This results in a DW width that is smaller than those calculated using of other definitions. This effect is particularly pronounced for the BPW, as it is characterized by a divergence of magnetization gradients at its core. 	Furthermore, note that our choice to consider a diameter of \SI{90}{nm} in the simulations, compared to the \SI{120}{nm} used in the experiments, is expected to nearly double the observed discrepancy as the wall width scales roughly with the square of the wire diameter for such large diameters \cite{bib-FRU2015b}.

\section{\label{sec:topology}{Topological analysis of BPW circulation switching}}

The switching mechanism described in the previous section involves the interplay between topological objects in the bulk and at the surface. In magnetism, vector field analysis using topological numbers \cite{Braun2012} usually applies to the objects evolving either exclusively in volumes or exclusively at surfaces, representative of thin and flat structures such as strips. To our knowledge, the mathematical generalization of the topological Hopf invariant formula \cite{Whitehead} is not yet available and its derivation is out of our scope here. Thus, given the absence of rigorous mathematical formalism dealing simultaneously with the volume and its bounding surface, we decided to describe our observations using topological numbers definitions applied to topological defects in magnetism in three and two dimensions separately \cite{Mermin1979}. Such classification of topological states allows to support the empirical rules reported in the previous section, which we expect should apply to any 3D nanomagnetic system.

The Bloch point is a true topological point defect (or singularity) consistent with the fact that the vector field order parameter $\vect{m}$ has dimension $d’=2$, lower than the spatial dimension $d=3$ \cite{bib-BRA2012}. To the contrary and as already discussed, the micromagnetic vortex (antivortex) at the wire surface is not a singularity, as $d’=2$ and $d=2$. In order to apply the established mathematical formalism \cite{Mermin1979} in the latter case to describe vortices and antivortices we consider the XY vector field consisting of the unit vectors of the two $d’=2$ magnetization components tangential to the wire surface, \ie, with $d’=1$ and $d=2$. In this gedanken (or fictitious) planar magnetization field, vortices and antivortices would be singularities. However, they are not in the full 3D vector field, so we label them simply with the equivalent integer topological number.

\begin{table}[t]
\begingroup
\setlength{\tabcolsep}{10pt}
\renewcommand{\arraystretch}{1.5}
    \centering
     \caption{Topological characteristics during the BPW circulation switching depicted (a)~in \subfigref{fig:BPW-switching}{b} for $\alpha=1$, and (b)~in \subfigref{fig:BPW-switching}{c} for $\alpha=0.1$. We use following notations: lower index in $()_1$ indicates the order of V/AV pair appearance in time. Red (orange) color corresponds to V(AV) of positive polarity, blue (green) corresponds to V(AV) of negative polarity. The labels (i)-(xii) correspond to particular events highlighted in \subfigref{fig:BPW-switching}{b-c}.}

    \begin{tabular}[t]{p{0.015\textwidth}||p{0.02\textwidth}c||p{0.16\textwidth}p{0.02\textwidth}}
             \hline\hline
        (a) &  $\omega_\mathrm{vol}$ & $C_{BPW}$&  $\sum (\omega_\mathrm{surf}\cdot p)_n$ & $=$ \\
     \hline
        i & -1 & C+ & -- & 0 \\
        ii & -1 & C+ & $\textcolor{blue}{(-1}\textcolor{teal}{+1)}_1$& 0 \\
        iii & 0 & -- & $\textcolor{red}{(+1}\textcolor{teal}{+1)}_1$ & 2\\
        iv & 0 & -- &$ \textcolor{red}{(+1}\textcolor{teal}{+1)}_1+\textcolor{blue}{(-1}\textcolor{teal}{+1)}_2$& 2\\
        v & 0 & -- & $\textcolor{red}{(+1)}_1+\textcolor{teal}{(+1)}_2$& 2\\
        vi & -1 & C- & --& 0 \\

        \hline\hline
            (b) &  $\omega_\mathrm{vol}$ & $C_{vol}$& $\sum (\omega_\mathrm{surf}\cdot p)_n$& $=$ \\
         \hline
            vii & -1 & C+ & -- & 0\\
            viii & -1 & C+ & $\textcolor{red}{(+1}\textcolor{orange}{-1)}_1$ & 0\\
            ix & -1 & C+ & $\textcolor{red}{(+1}\textcolor{orange}{-1)}_1$+$\textcolor{blue}{(-1}\textcolor{teal}{+1)}_2$ & 0\\
            x & 0 &--
             & \begin{tabular}{cc}$\textcolor{red}{(+1)}_1+\textcolor{blue}{(-1}\textcolor{teal}{+1)}_2+$\\$+\textcolor{red}{(+1}\textcolor{orange}{-1)}_3+\textcolor{teal}{(+1)}_4$ \end{tabular}& 2 \\
            xi & -2 & C- C- & $\textcolor{blue}{(-1)}_2+\textcolor{orange}{(-1)}_3$& -2 \\
            xii & -1 &  C- & -- & 0\\
            \hline\hline
    \end{tabular}
 \endgroup
    \label{tab:table-winding}
\end{table}

In the previous paragraph we explained the physical relevance of the topological numbers in 3d and 2d. In the following we present a rigorous way to implement their numerical identification in a magnetization field. For magnetic order parameter $\vect{m}=\{m^a,m^b,m^c\}$ the topological number of a point defect in volume is calculated using following surface integral \cite{Mermin1979}:
\begin{equation}
w =\frac{1}{8\pi} \int_{\partial V}\epsilon_{abc} m^{a} \mathrm{d} m^b \wedge \mathrm{d} m^c,
\label{eq:3d-winding}
\end{equation}
where $\wedge$ is the wedge product of differential forms on the $\partial V$ surface (also called 2-blade) and $\epsilon_{abc}$ is  Levi-Civita symbol. In short, the blade product of two vectors is the algebraic area of the parallelogram defined by the two vectors. \eqnref{eq:3d-winding} is a general form that can be useful when considering an arbitrary volume boundary, while the more common writing in cartesian coordinates $\{a,b,c\}=\{x,y,z\}$ is:
\begin{equation}
w_\mathrm{vol} = \int_{\partial V} [Q_x \mathrm{d} y \wedge \mathrm{d} z+Q_y \mathrm{d} z \wedge \mathrm{d} x+Q_z \mathrm{d} x \wedge \mathrm{d} y]\,,
\label{eq:3d-winding-cartesian}
\end{equation}
with $Q_x=\frac{1}{4\pi}\vect{m}\cdot(\partial_y \vect{m}\cross\partial_z\vect{m})$, $Q_y=-\frac{1}{4\pi}\vect{m}\cdot(\partial_x \vect{m}\cross\partial_z\vect{m})$, $Q_z=\frac{1}{4\pi}\vect{m}\cdot(\partial_x \vect{m}\cross\partial_y\vect{m})$ and $\vect{Q}=\{Q_x,Q_y,Q_z\}$ also known as vorticity vector \cite{bib-DON2020b}.  
In practical terms, the 2-blade product is analogous to the cross product of two vectors, it yields $w_\mathrm{vol}=+ 1$ for the tail-to-tail configuration and $w_\mathrm{vol}=- 1$ for the head-to-head configuration of the Bloch point, independently on circulation C$+$ or C$-$. For an in-plane (XY) magnetic order parameter $\vect{m}=\{m^a,m^b\}$ at a surface, the topological number is calculated using the following closed curve $\Gamma$ integral \cite{Mermin1979}
\begin{equation}
w_\mathrm{surf} =\frac{1}{2\pi} \int_{\Gamma} m^{a} \mathrm{d} m^b- m^b\mathrm{d} m^a,
\label{eq:2d-winding}
\end{equation}
this gives $w_\mathrm{surf}=+1$ for the vortex type configuration and $w_{surf}=-1$ for the antivortex type configuration. By definition for an XY model these values do not take into account  the vortex (antivortex) polarities $p$ of the full micromagnetic textures with respect to the external normal, which are related to the distribution of magnetic charges at the wire surface. Therefore, one needs to consider the quantity $w_\mathrm{surf}\cdot p$ to fully link surface and volume topological charges describing the $d’=2$ texture.

\tabref{tab:table-winding}a and \tabref{tab:table-winding}b summarize the topological numbers
computed at every time step denoted by a letter in  \figref{fig:BPW-switching}{b-c}, \ie, for $\alpha = 1$ and $\alpha = 0.1$, respectively.

We observe  that the quantity $\Delta \omega= \omega_\mathrm{vol}-\frac{1}{2}\sum(\omega_\mathrm{surf} \cdot p)_n$ is an invariant over any magnetization process. One may indeed come back to the set of volume/surface interaction rules listed in the previous section, and check the conservation of $\Delta \omega$ for each. The value of the invariant is $-1$, set by the initial conditions of a head-to-head Bloch point and no surface topological texture.  We may note that in many situations the topological charge is either entirely in the volume or at the surface, \ie, (i,ii,vi,vii,viii,ix,xii) and (iii,iv,v,x), respectively. However, there are situation in which a charge is found in both, here (xi). We  simulated the case of a tail-to-tail DW configuration, which also follows the conservation of $\Delta \omega$, in that case with value $+1$. This universality makes this rule powerful in the analysis of micromagnetics in 3D confined systems, enabling the description of transformations involving different topological objects, such as it is the case here. Note that this conservation rule is similar to the one described two decades ago in the context of flat and thin elements, linking topological objects at the surface and at its edges\cite{bib-TCH2005}.

\begin{figure}[t]
\centering\includegraphics[width=\linewidth]{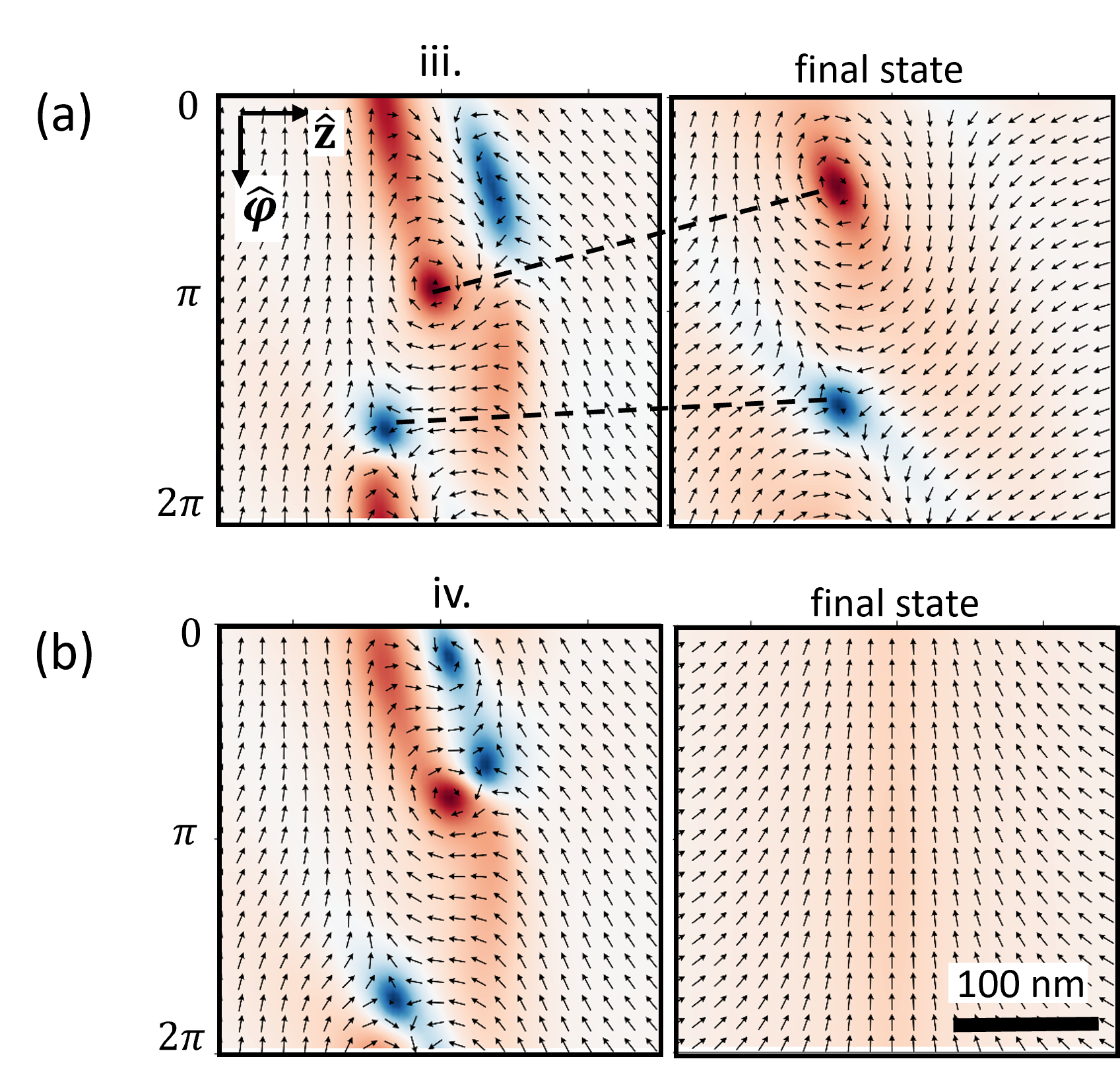}
\caption{Final state dependence on pulse length. The unrolled 2-dimensional ($\unitvectz$ - $\unitvectvarphi$) cylinder surfaces are viewed from the outside. The color map represents the radial magnetization component, where $m_{\rho}=1$ is red and $m_{\rho}=-1$ is blue.  (a) The final state after the current is switched off at event (iii) ($t= \SI{2.80}{ns}$) of \figref{fig:BPW-switching} shows V-AV stabilization, i.e., a TVW configuration. (b) The final state after the current is switched off at event (iv) ($t= \SI{2.98}{ns}$) of \figref{fig:BPW-switching} shows V-AV annihilation, i.e., a BPW configuration.}
\label{fig:TVW-stabilization}
\end{figure}

\section{\label{sec:TVW-transformation}Controlled conversion of domain wall type}

In their initial investigations, it was considered that BPWs and TVWs followed their own dynamics and could not change into one another due to their different topology \cite{bib-FOR2002,bib-HER2002a,bib-THI2006}, unless there was a change in the material or geometrical parameters, such as in a cone-shaped wire \cite{bib-HER2004}. However, this is only true for low diameter values, typically $\SI{50}{nm}$ or less, due to the prohibitive cost of exchange energy that would be involved during the transformation. The transformation from TVW to BPW was  first reported experimentally under an applied axial magnetic field \cite{bib-FRU2019}. The underlying microscopic process was explained by simulation, related to the differential azimuthal speed of surface V and AV leading to their collision. They merge and give rise to a Bloch Point as their polarity is opposite, ending up in a BPW in the relaxed state. Later on, the same transformation was reported to happen under an \OErsted field as the driving force \cite{bib-FRU2021}. However, the reverse transformation had not been predicted. In the following, we illustrate how to controllably induce this reverse transformation, from BPW to TVW, exploiting the non-Bloch Point nature of the transient state during the circulation switching process, hereby explaining the transformations reported in section \secref{sec-experiments}.

The dynamic states (iii), (iv) and (v) listed in \tabref{tab:table-winding}a and \subfigref{fig:BPW-switching}{b} are called topologically trivial in the volume, \ie, with $w_\mathrm{vol}=0$. As this is the case for a TVW, at this time we switched off the current in the simulation and let the system relax. \subfigref{fig:TVW-stabilization}{a-b} displays the resulting temporal evolution of V/AV pairs positions along the axial $\unitvectz$ and azimuthal $\unitvectvarphi$ directions when switching off the current at times (iii) and (iv), respectively.  Until the current is switched off, the plot is the same as in \subfigref{fig:BPW-switching}{b}. For the total pulse width of $\SI{2.80}{ns}$, the final state is a TVW. In contrast,  for a pulse width of $\SI{2.98}{ns}$, the final state is a BPW of reversed circulation. We attribute such difference to the number of neighboring surface topological textures. Indeed, initially, a single V/AV pair favors TVW configuration stabilization, whereas a high density of V/AV pairs on the surface promotes their annihilation with Bloch Point injection into the volume.

We conducted a series of simulations to generalize the results of \figref{fig:TVW-stabilization}, varying the pulse width and damping parameter. We find that in all cases, if the width of the pulse is in a suitable range, a pair of V/AV leads the dynamics and remains in the system in the final relaxed state without re-introducing a Bloch Point, ending up in a TVW state. If the pulse is too short, the final state is a BPW with the initial circulation, only slightly translated along the axial direction. Finally, if the pulse is too long so that a second V/AV pair has been nucleated, the dynamics continues and ends up in a BPW with reversed circulation. A deterministic condition for BPW-to-TVW transformation seems to be that one single V/AV pair exists in the system at the time the pulse is switched off.

\section{\label{sec:conclusion}Conclusion}

We have investigated the \OErsted-field-induced switching of circulation of Bloch-point walls (BPWs) in cylindrical nanowires, as a prototypical situation of a magnetization texture in a three-dimensional nano-object. Through micromagnetic simulations, we have predicted that the dynamical switching process involves the interplay of one or more Bloch points, which are bulk micromagnetic singularities, with one or more pairs of surface vortex/antivortex. We evaluate both a bulk topological number highlighting the number and type of Bloch points, and a surface topological number based on the mapping of magnetization onto a gedanken $XY$ tangential magnetization field. We exhibit a topological invariant linking bulk and surface topological numbers along with vortex/antivortex polarity. During any magnetization process, the topological charge may shift from volume to surface and vice-versa, defining different categories of topology for the magnetization textures. We show that the application of a nanosecond pulse of current with suitable duration can allow to select the topology of the final state, matching that of a transient state occurring at a specific time step. While demonstrate experimentally the effectiveness of this approach, converting Bloch-point walls into Transverse-Vortex walls. We expect that both the existence of a topological invariant and the possibility to select the topology the final state from that of a transient states apply to any 3D nanomagnetic system. This should provide guidelines to categorize and control the many topologies and dynamic control of magnetization textures offered by 3D nanomagnetism, with potential application in spintronics and magnetic memory devices for 3D electronics.

The data that support the findings of this article are openly available \cite{zenodoAlvaroGomez2025}.

\section{\label{sec:Acknowledgments}Acknowledgments}

This work received financial support from the French RENATECH network, implemented at the Upstream Technological Platform in Grenoble PTA  (ANR-22-PEEL-0015), from the French National Research Agency (Grant No. ANR-17-CE24-0017 MATEMAC-3D; Grant ANR-22-CE24-0023 DIWINA) from the Spanish MCIN/AEI/10.13039/501100011033 through Projects PID2020-117024GB-C43, TED2021-130957B-C52 and CEX2020- 001039-S).
We also acknowledge support from the Regional Government of Madrid under project TEC-2024/TEC-380 Mag4TIC-CM.  S. R-G acknowledges support from the Humboldt foundation grant 1223621 and Marie Curie fellowship grant GAP-101061612. We acknowledge support from the team of the Nanofab platform (CNRS Néel institut) and from the ALBA in-house research program and MISTRAL beamline.

\bibliography{Fruche8-2,Lauracomp-2} 

\end{document}